\documentclass[preprint,amsmath,eqsecnum,nofootinbib,tightenlines]{revtex4}
\usepackage{graphicx}
\def\Slash#1{{\ooalign{\hfil$#1$\hfil\crcr\hfil$/$\hfil}}}

\begin{document}
\baselineskip=16pt
\begin{flushright}
\begin{minipage}{22mm}{\small%
UT-HET 036\\%
KUNS-2263}
\end{minipage}
\end{flushright}
\title{\vspace*{13mm}%
Seesaw Neutrino Signals at the Large Hadron Collider\vspace*{5mm}}
\author{%
Shigeki Matsumoto,$^{\!a}$ 
Takehiro Nabeshima,$^{\!a}$
and Koichi Yoshioka$^b$
}
\affiliation{%
$^a$Department of Physics, University of Toyama, Toyama 930-8555, Japan\\
$^b$Department of Physics, Kyoto University, Kyoto 606-8502, Japan%
\vspace*{15mm}
}
\begin{abstract}\vspace*{3mm}\noindent%
We discuss the scenario with gauge singlet fermions (right-handed
neutrinos) accessible at the energy of the Large Hadron Collider. The
singlet fermions generate tiny neutrino masses via the seesaw
mechanism and also have sizable couplings to the standard-model
particles. We demonstrate that these two facts, which are naively not
satisfied simultaneously, are reconciled in the five-dimensional
framework in various fashions, which make the seesaw mechanism
observable. The collider signal of tri-lepton final states with
transverse missing energy is investigated for two explicit examples of
the observable seesaw, taking account of three types of neutrino mass
spectrum and the constraint from lepton flavor violation. We find by
showing the significance of signal discovery that the collider
experiment has a potential to find signals of extra dimensions and the
origin of small neutrino masses.
\end{abstract}
\maketitle
\thispagestyle{empty}\setcounter{page}{0}
\newpage

%%%%%%%%%%%%%%%%%%%%%%%%%%%%%%%%%%%%%%%%%%%%%%%%%%%%%%%%%%%%%%%%%
{\centering\section{Introduction}}

The neutrino property, in particular its tiny mass scale is one of the
most important experimental clues to find new physics beyond the
standard model~(SM)~\cite{neu_review}. From a theoretical viewpoint,
the seesaw mechanism~\cite{seesaw} has been known to naturally induce
small neutrino masses by integrating out new heavy particles
(right-handed neutrinos) which interact with the left-handed SM
neutrinos. In this Type I seesaw scheme, the right-handed neutrinos
have intermediate-scale masses for 
obtaining ${\cal O}(\text{eV})$ light neutrinos. The seesaw effect 
appears as higher-dimensional operators suppressed by their heavy mass
scale and are usually negligible in low-energy effective
theory. Alternatively, TeV-scale right-handed neutrinos are found from
the seesaw formula to have very weak couplings to the SM sector and
their signals would not be naively detected in future experiments such
as the CERN Large Hadron Collider (LHC).

It seems therefore difficult to directly observe heavy states which
are relevant to suppressing neutrino masses. In the previous
work~\cite{HMY}, it was pointed out that an observable seesaw
mechanism can be implemented in a five-dimensional framework where all
the SM fields are confined in a four-dimensional boundary while
right-handed neutrinos propagate in the bulk of extra-dimensional
space~\cite{DDG,ADDM}. The existence of extra dimensions is also one
of the exciting candidates for new physics beyond the
SM~\cite{ExD_review} and related neutrino phenomenology has been 
extensively studied in the literature~\cite{neuExD}. In the above
framework, the right-handed neutrinos and their extra-dimensional
partners exist around the TeV scale and have sizable SM gauge and
Yukawa couplings in the low-energy effective theory, while the
seesaw-induced masses are made small.

The SM neutrinos have tiny masses due to a slight violation of the
lepton number. This fact implies that the events with same-sign
di-lepton final states~\cite{dileptons} may be too rare to be observed
unless, e.g., some particular flavor structure is assumed in neutrino
mass matrices. In this paper, as in our previous work, we analyze
lepton number conserving processes, in particular, the tri-lepton
signal with large missing transverse 
energy: $pp\to\ell^\pm\ell^\mp\ell^\pm\nu(\bar\nu)$. This process is
expected to be effectively detected at the LHC because only a small
fraction of SM processes contributes to the background against the
signal. The LHC signatures are studied in typical two types of
observable seesaw models in five dimensions and with three types of
neutrino mass patterns allowed by the present experimental
data~\cite{neu_analysis}.

We also present various extra-dimensional approaches which provide the
situation that TeV-scale right-handed neutrinos generate a proper
scale of seesaw-induced masses and simultaneously have observable
interactions to the SM fields. They include boundary Majorana mass
terms, boundary conditions for bulk neutrinos, the 
AdS$_5$ gravitational background, and their combinations. These
scenarios do not rely on particular (singular or aligned) generation
structure of mass matrices, and is available even in the
one-generation case. For such TeV-scale particles with sizable
couplings to the SM sector, the collider experiment will generally
have a potential to find a signal of extra dimensions and the origin
of small neutrino masses.

This paper is organized as follows. In Section~\ref{sec:framework}, we
formulate the general five-dimensional setup for neutrino physics,
discussing the seesaw operation and the electroweak
Lagrangian. Several explicit models for the observable seesaw are
presented for the collider study. In Section~\ref{sec:LHC}, after the
discussion of phenomenological constraints and representative points
of model parameters, we numerically investigate the LHC signatures of
the seesaw models given in Section~\ref{sec:framework} and illustrate
the significance for the signal discovery. In
Section~\ref{sec:others}, we further show that various different
configurations for the observable seesaw are viable even in one extra
dimension, giving the low-energy effective vertices of heavy
neutrino fields. Section~\ref{sec:summary} is devoted to summarizing
our results and discussing future work.

\bigskip\bigskip\bigskip

%%%%%%%%%%%%%%%%%%%%%%%%%%%%%%%%%%%%%%%%%%%%%%%%%%%%%%%%%%%%%%%%%
{\centering\section{General framework}%
\label{sec:framework}}

%%%%%%%%%%%%%%%%%%%%%%%%%%%%%%%%%%%%%%%%%%%%%%%%%%%%%%%%%%%%%%%%%
{\centering\subsection{Five-dimensional Setup}}

Let us consider a five-dimensional theory where the extra space is
compactified on the $S^1/Z_2$ orbifold with the radius $R$. The
SM fields are confined on the four-dimensional boundary
at $x^5=0$. Besides the gravity, only SM gauge singlets can propagate
in the bulk not to violate the charge
conservation~\cite{DDG,ADDM}. The gauge-singlet Dirac 
fermions $\Psi_i$ ($i=1,2,3$) are introduced in the bulk which
contain right-handed neutrinos and their chiral partners. The 
Lagrangian up to the quadratic order of spinor fields is given by
\begin{eqnarray}
  e^{-1}{\cal L} \;=\; i\overline\Psi D\hspace{-2.5mm}/\,\Psi
  -\overline{\Psi}(m_d+im_{d5}\gamma_5)\theta(x^5)\Psi
  -\frac{1}{2}\big[\,\overline{\Psi^c}
  (M+M_5\gamma_5)\Psi+\text{h.c.}\big].
\end{eqnarray}
The conjugated spinor is defined 
as $\Psi^c=i\gamma^2\gamma^0\gamma_5\overline{\Psi}{}^{\rm t}$ such
that it is Lorentz covariant in five dimensions. The covariant
derivative generally contains the contribution of spin connection
given by the f\"unfvein. The bulk Dirac mass term involves the step 
function $\theta(x^5)$ so that it is invariant under 
the $Z_2$ reflection. Such an odd-function dependence could originate
from some field expectation value. The bulk mass 
parameters $m_d$, $m_{d5}$, $M$ and $M_5$ are $Z_2$-parity even and
generally depend on the extra-dimensional coordinate $x^5$ which comes
from the delta-function dependence (resulting in localized mass terms)
and/or the background geometry such as the warp factor in AdS$_5$. We
also introduce the mass terms between bulk and boundary fields:
\begin{eqnarray}
  {\cal L}_m \;=\; -\big(\overline{\Psi} mL+
  \overline{\Psi^c}m^cL\big)\delta(x^5) +{\rm h.c.},
  \label{boundary}
\end{eqnarray}
where $m$ and $m^c$ denote the mass parameters after the electroweak
symmetry breaking (the original Yukawa term will be given
later). Throughout this paper, we take the fundamental scale of
five-dimensional theory as the unit of mass dimension-ful
parameters. The boundary spinors $L_i$ ($i=1,2,3$) contain the
left-handed neutrinos $\nu_i$. The $Z_2$ parity implies that either
component in a Dirac fermion $\Psi$ vanishes at the boundary ($x^5=0$)
and therefore either of $m$ and $m^c$ becomes irrelevant.\footnote{The
exception is the generation-dependent parity assignment on bulk
fields~\cite{parity}. We do not consider such a possibility in this
paper.} In the following we assign the even $Z_2$ parity to the upper
(right-handed) component of bulk fermions
\begin{eqnarray}
  \Psi(-x^5) \;=\; \gamma_5\Psi(x^5),
\end{eqnarray}
and will drop the $m^c$ term. In the above, while we only consider the
boundary terms at $x^5=0$, other boundary terms at $x^5=\pi R$ can
also be written down in the same fashion and have physical relevance
on curved backgrounds and/or with complicated field configurations.

With a set of boundary conditions, the bulk fermions $\Psi_i$ are
expanded by Kaluza-Klein (KK) modes with their kinetic terms being
properly normalized
\begin{eqnarray}
  \Psi(x,x^5) \;=\; \left(\begin{array}{l}
    \sum\limits_n \chi^n_R(x^5)N_R^n(x) \\[3.4mm]
    \sum\limits_n \chi^n_L(x^5)N_L^n(x)
  \end{array} \right).
\end{eqnarray}
The wavefunctions $\chi_{R,L}^n$ are generally matrix-valued in the
generation space and we have omitted the generation indices for
notational simplicity. After integrating over the fifth dimension, we 
obtain the neutrino mass matrix in four-dimensional effective theory:
\begin{eqnarray}
  {\cal L}_4 \;=\; i{\cal N}^\dagger\sigma^\mu\partial_\mu{\cal N}
  -\frac{1}{2}\big(\,{\cal N}^{\rm T}\epsilon
  {\cal M}{\cal N}+{\rm h.c.}\,\big),
  \label{L4}
\end{eqnarray}
where $\epsilon=i\sigma^2$, and ${\cal N}$ is composed of the boundary
neutrinos and the KK modes ${\cal N}=
(\nu\,|\,\epsilon N_R^{\,0\,*},\epsilon N_R^{\,1\,*},N_L^{\,1},
\epsilon N_R^{\,2\,*},N_L^{\,2},\cdots)\equiv(\nu\,|\,N)$. The zero
modes of the left-handed components have been extracted according to
the boundary condition. The neutrino mass matrix for $(\nu\,|\,N)$ is
given by
\begin{eqnarray}
\renewcommand{\arraystretch}{1.2}
{\cal M} \;=\; \left(%
\begin{array}{c|cccc}
  & \,m_0^t & \,m_1^{\rm t} & 0 & \cdots \\ \hline
  m_0 & -M_{R_{00}}^* & -M_{R_{01}}^* & M_{K_{01}}^{} 
  & \cdots \\[1mm]
  m_1 & -M_{R_{10}}^* & -M_{R_{11}}^* & M_{K_{11}}^{} & \cdots \\[1mm]
  0 & \phantom{-}M_{K_{10}}^{\rm t} 
  & \phantom{-}M_{K_{11}}^{\rm t} & M_{L_{11}}^{} & \cdots \\[1mm]
  \vdots & \vdots & \vdots & \vdots & \ddots
\end{array}\right)
\;\equiv\; \left(
\begin{array}{c|ccc}
  & & M_D^{\rm t} & \\ \hline
  & & & \\
  \!\!M_D & ~ & M_N & \\
  & & &
\end{array}\right),
\end{eqnarray}
where the Majorana masses $M_{L,R}$, the KK masses $M_K$, and the
boundary Dirac masses $m_n$ are
\begin{alignat}{2}
  M_{R_{mn}} &=\; \int_{-\pi R}^{\pi R}\!dx^5\,
  (\chi^m_R)^{\rm t} (M+M_5) \chi^n_R,  &\qquad
  M_{K_{mn}} &= \int_{-\pi R}^{\pi R}\!dx^5\, 
  (\chi^m_R)^\dagger(-\omega\partial_5+m_d+im_{d5})\chi^n_L,
  \nonumber \\[1mm]
  M_{L_{mn}} &=\; \int_{-\pi R}^{\pi R}\!dx^5\,
  (\chi^m_L)^{\rm t} (M-M_5) \chi^n_L,  &\qquad
  m_n \;&=\; \chi^n_R\!{}^\dagger(0)m.
\end{alignat}
In the expression of KK masses $M_K$, the factor $\omega$ is related to
the five-dimensional geometry, for example, $\omega=1$ for the flat
background and $\omega=e^{-k|x^5|}$ for the AdS$_5$ background with
the curvature $k$. It is noticed that $M_{K_{mn}}$ becomes
proportional to $\delta_{mn}$ if $\chi_{R,L}^n$ are the eigenfunctions
of the bulk equations of motion, and $M_{R,L_{mn}}$ also becomes
proportional to $\delta_{mn}$ if the bulk mass 
parameters $M$, $M_5$ are independent of the coordinate $x^5$.

\bigskip\bigskip\bigskip

%%%%%%%%%%%%%%%%%%%%%%%%%%%%%%%%%%%%%%%%%%%%%%%%%%%%%%%%%%%%%%%%%
{\centering\subsection{Seesaw and Electroweak Lagrangian}}

We further implement the seesaw operation 
assuming ${\cal O}(m_n)\ll{\cal O}(M_K),\,{\cal O}(M_{R,L})$ and
find the induced Majorana mass matrix for three-generations light
neutrinos\footnote{In theory with more than one extra dimensions, this
matrix product (the sum of infinite KK modes) generally diverse
without some regularization~\cite{bf}.}
\begin{eqnarray}
  M_\nu \;=\; -M_D^{\text{t}}M_N^{-1}M_D^{}.
\end{eqnarray}
It is useful for later discussion of collider phenomenology to write
down the electroweak Lagrangian in the basis where all the mass
matrices are generation diagonalized. The original Lagrangian of
four-dimensional neutrinos comes from \eqref{L4} and the SM part. The
kinetic and mass terms and the interactions to the electroweak gauge
bosons are given in the mass eigenstate basis $(\nu_d,N_d)$ as follows:
\begin{eqnarray}
  {\cal L}_{\rm EW} &=& i \nu_d^{\,\dagger}\sigma^\mu\partial_\mu\nu_d
  + iN_d^{\,\dagger}\sigma^\mu\partial_\mu N_d \,
  -\frac{1}{2}\big(\nu_d^{\,\rm t}\epsilon M_\nu^d\,\nu_d
  +N_d^{\,\rm t}\epsilon M_N^d\,N_d
  +{\rm h.c.}\big)   \nonumber \\[1mm] 
  && \qquad 
  +\frac{g}{\sqrt{2}}\Big[W_\mu^\dagger e^\dagger\sigma^\mu
  U_{\rm MNS} \big(\nu_d+VN_d\big) +\text{h.c.}\Big]  \nonumber \\[1mm]
  && \qquad\qquad
  +\frac{g}{2\cos\theta_W}Z_\mu\big(\nu_d^\dagger+
  N^{\,\dagger}_dV^\dagger\big)
  \sigma^\mu\big(\nu_d+V N_d\big), \nonumber
\end{eqnarray}
where $W_\mu$ and $Z_\mu$ are the electroweak gauge bosons and $g$ is
the $SU(2)_{\rm weak}$ gauge coupling constant. The 2-component
spinors $\nu_d$ are three light neutrinos for which the seesaw-induced 
mass matrix $M_\nu$ is diagonalized
\begin{eqnarray}
  M_\nu^d \;=\; U_\nu^{\rm t}\,M_\nu\,U_\nu,  \qquad
  U_\nu\,\nu_d \;=\; \nu-M_D^\dagger M_N^{-1\,*}N,
\end{eqnarray}
and $N_d$ denote the infinite number of neutrino KK modes for which
the bulk Majorana mass matrix $M_N$ is diagonalized both in the
generation and KK spaces by a unitary matrix $U_N\,$:
\begin{eqnarray}
  M_N^d \,=\; U_N^{\rm t}\,M_N^d\,U_N,  \qquad
  U_N N_d\ \,=\; N+M_N^{-1}M_D^{}\,\nu.
\end{eqnarray}
The lepton mixing matrix measured in the neutrino oscillation
experiments is given by $U_{\rm MNS}=U_e^\dagger U_\nu$ where $U_e$ is
the left-handed rotation matrix for diagonalizing the charged-lepton
Dirac masses. It is interesting to find that the model-dependent parts
of electroweak gauge vertices are governed by a single 
matrix $V$ which is defined as
\begin{eqnarray}
  V \;=\; U_\nu^\dagger M_D^\dagger M_N^{-1\,*}U_N.
\end{eqnarray}
When one works in the basis where the charged-lepton sector is flavor
diagonalized, $U_\nu$ is fixed by the neutrino oscillation matrix.

The neutrinos also have the interaction to the electroweak doublet
Higgs $H$. If assuming $H$ lives in the four-dimensional fixed point
at $x^5=0$, the boundary Dirac mass~\eqref{boundary} comes from the
Yukawa coupling
\begin{eqnarray}
  {\cal L}_h \;=\; -y\tilde H^\dagger\overline{\Psi}L\delta(x^5)
  +\text{h.c.},
\end{eqnarray}
where $\tilde H=\epsilon H^*$. The doublet Higgs $H$ has a
non-vanishing expectation value $v/\sqrt{2}$ and its 
fluctuation $h(x)$. After integrating out the fifth dimension and
diagonalizing mass matrices, we have
\begin{eqnarray}
  {\cal L}_h \;=\; \frac{-h}{v}\sum_n
  \big[(N_d^{\rm t}-\nu_d^{\rm t}V^*)U_N^{\rm t}\big]_{R_n}\!\!
  m_nU_\nu\,\epsilon(\nu_d+VN_d) +\text{h.c.},
\end{eqnarray}
where $[\cdots]_{R_n}$ means the $n$-th mode of the right-handed
component.

\bigskip\bigskip\bigskip

%%%%%%%%%%%%%%%%%%%%%%%%%%%%%%%%%%%%%%%%%%%%%%%%%%%%%%%%%%%%%%%%%
{\centering\subsection{Models for Observable Seesaw}}

The heavy neutrino interactions to the SM fields are determined by the
mixing matrix $V$ both in the gauge and Higgs 
vertices. The $3\times\infty$ matrix $V$ is determined by the matrix
forms of neutrino masses in the original 
Lagrangian ${\cal L}+{\cal L}_m$. The matrix elements in $V$ have
the experimental upper bounds from electroweak physics, as will be
seen later. Another important constraint on $V$ comes from the
low-energy neutrino experiments, namely, the seesaw-induced masses
should be of the order of eV scale, which in turn specifies the scale
of heavy neutrino masses $M_N$. This can be seen from the definition
of $V$ by rewriting it with the light and heavy neutrino mass eigenvalues
\begin{eqnarray}
  V \;=\; i(M_\nu^d)^{\frac{1}{2}}P(M_N^d)^{-\frac{1}{2}},
  \label{V}
\end{eqnarray}
where $P$ is an arbitrary $3\times\infty$ matrix 
with $PP^{\rm t}=1$. Therefore one naively expects that, with a fixed
order of $M_\nu^d\sim10^{-1}\,\text{eV}$ and $|V|\gtrsim10^{-2}$ for
the discovery of experimental signatures of heavy neutrinos, their
masses should be very light and satisfy $M_N^d\lesssim$~keV (this does
not necessarily mean the seesaw operation is not justified 
as $M_\nu^d$ is fixed). The previous collider studies on TeV-scale
right-handed neutrinos~\cite{TeVRH} did not impose the seesaw
relation~\eqref{V} and have to rely on some assumptions for
suppressing the necessarily induced masses $M_\nu$. For example, the
neutrino mass matrix has some singular generation structure, otherwise
it leads to the decoupling of seesaw neutrinos from collider physics.

We here present two scenarios in which heavy neutrino modes are
accessible at future colliders. The numerical study of these two
models will be performed in the next section. It is noted that they
are illustrative examples and there are many other possibilities for
the observable seesaw with extra dimensions. We will comment on such
various alternatives in a later section.

\bigskip\bigskip\bigskip

%%%%%%%%%%%%%%%%%%%%%%%%%%%%%%%%%%%%%%%%%%%%%%%%%%%%%%%%%%%%%%%%%
{\centering\subsubsection{Model 1 ~~$-$ Particular Majorana Masses $-$}%
\label{sec:Model1}}

A possible scenario for observable heavy neutrinos is to take a
specific value of Majorana mass parameters. Let us consider the
situation that the bulk Majorana mass $M$ and bulk Dirac masses are
vanishing on the Minkowski background. The Lagrangian is
\begin{eqnarray}
  {\cal L} \;=\; i\overline\Psi\partial\hspace{-2.2mm}/\,\Psi
  -\Big[\,\frac{1}{2}\overline{\Psi^c}M_5\gamma_5\Psi
  +\overline{\Psi} mL\delta(x^5) +\text{h.c.}\Big].
\end{eqnarray}
The equations of motion without bulk masses are solved by simple
oscillators and the mass matrices in four-dimensional effective 
theory \eqref{L4} are found
\begin{gather}
  \quad
  M_{K_{mn}} \,=\; -\frac{n}{R}\delta_{mn}, \quad\;\;
  M_{R_{mn}} \,=\; -M_{L_{mn}} \,=\; M_5\delta_{mn}, \quad\;\;
  m_n \;=\; \frac{m}{\sqrt{2^{\delta_{n0}}\pi R}}\,.
\end{gather}From these,
we find the seesaw-induced mass matrix and the mixing with heavy modes:
\begin{eqnarray}
  M_\nu &=& \frac{1}{2\pi R}\,m^{\rm t}
  \frac{\pi R|M_5|}{\tan(\pi R|M_5|)}\, \frac{1}{M_5^*}\,m,  \\[1mm]
  \nu \,&=& U_\nu\nu_d \,-\frac{\,m^\dagger}{\sqrt{2\pi R}}\,
  \bigg[ \frac{1}{M_5}\,\epsilon N_R^{\,0\,*}
  +\sum_{n=1}\frac{\sqrt{2}}{|M_5|^2\!-(n/R)^2}
  \Big(M_5^*\epsilon N_R^{\,n\,*}-
  \frac{n}{R}\,N_L^{\,n}\Big)\bigg]. \quad
\end{eqnarray}
The KK neutrinos have the mass 
eigenvalues $|M_5|$ and $\frac{n}{R}\pm|M_5|$ ($n\geq1$). The effect of
infinitely many numbers of KK neutrinos appears as the additional
factor $\pi R|M_5|/\tan(\pi R|M_5|)$. An interesting case is that (the
eigenvalues of) $M_5$ takes a specific 
value $|M_5|=\alpha/R$ where $\alpha$ contains half
integers~\cite{DDG}: the seesaw-induced mass $M_\nu$ is then
suppressed by the tangent factor (not only by large Majorana mass),
on the other hand, the heavy mode vertex $V$ is un-suppressed. This
fact realizes the situation that right-handed neutrinos in the seesaw
mechanism are observable at sizable rates in future collider
experiments~\cite{HMY} (see also \cite{BMOZ}).

As an explicit example, we consider flavor-independent Majorana 
masses $M_5=\frac{1}{2R}-\delta_M$ where $\delta_M$ is 
small ($\ll1/R$) and denotes a deviation from massless neutrinos. 
A vanishing $\delta_M$ makes the light neutrinos exactly massless,
where a complete cancellation occurs within the seesaw effects of
heavy neutrinos. As we will see, the parameter $\delta_M$ takes a tiny 
value for giving the correct neutrino mass scale.\footnote{That seems
a fine tuning of model parameters; the bulk Majorana masses must be
fixed almost exactly. This tuning is ameliorated by considering a
different extra-dimensional setup with the same neutrino mass matrix
(see Section~\ref{sec:b_cond}).} In the KK-mode picture, the mass
spectrum becomes almost vector-like and no chiral zero mode
exists. The seesaw-induced mass and the KK Dirac masses $M_n$ are
given by
\begin{eqnarray}
  M_\nu \;=\; \frac{\pi R\delta_M}{2}\,m^{\rm t}m, \qquad\quad
  M_n \;\simeq\; \frac{n-\tfrac{1}{2}}{R} \qquad (n\geq1).
\end{eqnarray}
We will consider $M_n\sim1/R={\cal O}(10^{2-3})\;\text{GeV}$ for
the LHC analysis of low-lying KK neutrinos. The neutrino Yukawa
coupling $y_\nu$ is expressed as
\begin{eqnarray}
  y_\nu \;=\; \frac{2}{\pi Rv}(\delta_M)^\frac{-1}{2}O^\dagger
  (M_\nu^d)^\frac{1}{2}U^\dagger_{\rm MNS}, 
\end{eqnarray}
where $O$ is an arbitrary $3\times3$ orthogonal matrix which generally
comes in reconstructing high-energy quantities from the low-energy
neutrino observables~\cite{CI}. That corresponds to the 
matrix $P$ in~\eqref{V}. 

\bigskip\bigskip\bigskip

%%%%%%%%%%%%%%%%%%%%%%%%%%%%%%%%%%%%%%%%%%%%%%%%%%%%%%%%%%%%%%%%%
{\centering\subsubsection{Model 2 ~~$-$ Light Dirac Neutrinos $-$}%
\label{sec:Model2}}

Another example of observable heavy states is realized by assuming
no Majorana mass for bulk neutrinos, which leads to lepton
number conservation while having sizable couplings to the SM
neutrinos. The Lagrangian is
\begin{eqnarray}
  {\cal L} \;=\; i\overline\Psi\partial\hspace{-2.2mm}/\,\Psi
  -\overline{\Psi}m_d\theta(x^5)\Psi
  -\big[\overline{\Psi} mL\delta(x^5) +\text{h.c.}\big].
\end{eqnarray}
The solution to the bulk equations of motion in the presence of bulk
Dirac masses are given by
\begin{gather}
\begin{align}
  \chi^0_R \;&=\; \frac{1}{\sqrt{\pi R}}\,f_0\,e^{-m_d|x^5|}, \\
  \chi^n_R \;&=\; \frac{1}{\sqrt{\pi R}}
  \Big[f_n\cos\Big(\frac{n}{R}x^5\Big)+
  \sqrt{1-f_n^2}\,\theta(x^5)\sin\Big(\frac{n}{R}x^5\Big)\,\Big], \\
  \chi^n_L \;&=\; \frac{1}{\sqrt{\pi R}}\sin\Big(\frac{n}{R}x^5\Big),
\end{align}\\[3mm]
  f_0 \;=\; \sqrt{\frac{\pi Rm_d}{1-e^{-2\pi Rm_d}}}, \qquad
  f_n \;=\; \frac{-n/R}{\sqrt{(n/R)^2+m_d^2}} \qquad  (n\geq1).
\end{gather}
The zero mode $N_R^{\,0}$ is massless at this stage and has a
localized wavefunction controlled by the bulk Dirac 
mass $m_d$. The $n$-th excited modes have the squared mass 
eigenvalues $(n/R)^2+m_d^2$. The mass matrices in four-dimensional
effective theory \eqref{L4} are found
\begin{gather}
  \quad
  M_{K_{mn}} \,=\; \sqrt{(n/R)^2+m_d^2\,}\,\delta_{mn}, \qquad
  M_{R_{mn}} \,=\; M_{L_{mn}} \,=\; 0, \qquad
  m_n \;=\; \frac{f_n}{\sqrt{\pi R}}\,m\,.
\end{gather}
While the excited modes are heavy ($\gtrsim1/R$), the zero mode has
no contribution from bulk and KK masses. Therefore the zero modes
compose of Dirac particles with the SM neutrinos and obtain their
masses from the SM Higgs field: ${\cal L}=
-m_0N_R^0{}^\dagger\nu+\text{h.c.}$. On the other hand, since the
excited modes have KK Dirac masses and no lepton number violation,
they do not give rise to the seesaw-induced mass $M_\nu$ (i.e.\ the
contributions from $N_R^{\,n}$ and $N_L^{\,n}$ are cancelled to each
other) and the right-handed components $N_R^{\,n}$ do not mix with the
left-handed SM neutrinos. We thus find the light Dirac neutrino masses
and the mixing with heavy modes:
\begin{eqnarray}
  m_0 &=& \sqrt{\frac{m_d}{1-e^{-2\pi Rm_d}}}\,m, \\
  \nu \,&=& U_\nu\nu_d \,-\frac{\,m^\dagger}{\sqrt{\pi R}}\,
  \sum_{n=1}\frac{n/R}{(n/R)^2+m_d^2}\,N_L^{\,n}.
\end{eqnarray}
The Dirac neutrino mass $m_0$ can be suppressed by the exponential
wavefunction factor $f_0$, while the heavy KK modes are kept
observable. For example, 
if $-Rm_d\sim8$, the ${\cal O}(\text{eV})$ neutrinos are obtained for
other parameters being on TeV scale. A negative value of $m_d$ means
that the zero mode is localized away from the SM boundary ($x^5=0$),
which situation leads to the suppression of Dirac neutrino 
mass $m_0$. The heavy modes have rather broad wavefunctions in the
bulk and the couplings to the SM sector are independent of the
exponential suppression. That allows the heavy modes to take
sizable boundary couplings and to be observed.

In this model, the light and KK neutrinos are all Dirac particles and
their mass eigenvalues are given by
\begin{eqnarray}
  M_\nu \;(\,=m_0)\;=\; \sqrt{\frac{m_d}{1-e^{-2\pi Rm_d}}}\,m, 
  \qquad\;
  M_n \;=\; \sqrt{(n/R)^2+m_d^2\,} \qquad (n\geq1).
\end{eqnarray}
We will consider $M_n\sim m_d={\cal O}(10^{2-3})\;\text{GeV}$ for
the LHC analysis of low-lying KK neutrinos. The neutrino Yukawa
coupling $y_\nu$ is expressed as
\begin{eqnarray}
  y_\nu \;=\; \frac{2}{v}\sqrt{\frac{1-e^{-2\pi Rm_d}}{2m_d}}\,
  U_RM_\nu^d U^\dagger_{\rm MNS},
\end{eqnarray}
where $U_R$ is the $3\times3$ unitary matrix which rotates the
three-generation right-handed zero modes so that the light Dirac mass
matrix $M_\nu$ is diagonalized to $M_\nu^d$.

\bigskip\bigskip\bigskip

%%%%%%%%%%%%%%%%%%%%%%%%%%%%%%%%%%%%%%%%%%%%%%%%%%%%%%%%%%%%%%%%%
{\centering\subsubsection{Model 3 ~~ $-$  Small Lepton Number Violation $-$}}

A slightly different model for observable heavy neutrinos is constructed 
by introducing small bulk Majorana mass into Model~2, which means the
light neutrinos are Majorana particles. The Lagrangian is
\begin{eqnarray}
  {\cal L} \;=\; i\overline\Psi\partial\hspace{-2.2mm}/\,\Psi
  -\overline{\Psi}m_d\theta(x^5)\Psi
  -\Big[\,\frac{1}{2}\overline{\Psi^c}M\Psi
  +\overline{\Psi} mL\delta(x^5) +\text{h.c.}\Big].
\end{eqnarray}
With non-vanishing Majorana masses, the lepton number is broken and
the seesaw-induced mass $M_\nu$ is generated by the integration of
heavy modes as in Model~1;
\begin{eqnarray}
  M_\nu &=& \frac{1}{2\pi R}\,m^{\rm t} \bigg[ \pi Rm_d 
  +\frac{\pi R\sqrt{|M|^2+m_d^2\,}}{\tanh\!\big(\pi R\sqrt{|M|^2+
  m_d^2\,}\big)}\bigg]\frac{1}{M^*}\,m.
\end{eqnarray}
The mixing with KK neutrinos has a similar expression to Model~2;
\begin{eqnarray}
  \nu \;\simeq\; U_\nu\nu_d \,-\frac{\,m^\dagger}{\sqrt{\pi R}}\,
  \bigg[\frac{f_0}{\sqrt{2}M}\,\epsilon N_R^{\,0\,*} 
  +\sum_{n=1}\frac{n/R}{(n/R)^2+m_d^2}
  \,N_L^{\,n}\bigg],
\end{eqnarray}
where we have assumed $|M|\ll 1/R,\,|m_d|$. The zero-mode
contribution is suppressed if it is enough separated from the SM
boundary with the localizing wavefunction, which 
implies $-Rm_d\gtrsim6$. It is found from the above expressions that
the seesaw neutrino mass and the couplings of heavy modes can be
determined independently that makes the seesaw mechanism
observable. The neutrino mass, i.e.\ the size of lepton number
violation is controlled by the bulk Majorana mass $M$. For example,  
if $-Rm_d\sim6$ and a few~\% mixing of heavy mode,
\begin{eqnarray}
  M \;\sim\; 10^3 \;\text{eV},
\end{eqnarray}
for eV seesaw-induced masses. The fundamental scale of
five-dimensional theory is irrelevant for this evaluation. The zero
mode $N_R^{\,0}$ obtains a Majorana mass of the order of $M$ and is
a light isolated particle with a negligible interaction to the SM sector.

In the end, the low-energy theory contains light Majorana neutrinos
with the seesaw-induced mass $M_\nu$, almost decoupled zero modes with
mass around keV scale, and heavy KK Dirac neutrinos. The mass
eigenvalues of these states are explicitly given by
\begin{eqnarray}
  M_\nu \;\simeq\; m^{\rm t}\bigg(\frac{M}{4m_d}\bigg)\,m, \qquad\;
  M_0 \;=\; M, \qquad\;
  M_n \;\simeq\; \sqrt{(n/R)^2+m_d^2\,} \quad\; (n\geq1).
\end{eqnarray}
The low-lying KK states would be observable at colliders 
for $M_n\sim m_d={\cal O}(10^{2-3})\;\text{GeV}$\@. The neutrino
Yukawa coupling has a similar expression to that in Model~1.

\bigskip\bigskip\bigskip

%%%%%%%%%%%%%%%%%%%%%%%%%%%%%%%%%%%%%%%%%%%%%%%%%%%%%%%%%%%%%%%%%
{\centering\section{Seesaw Signatures at the LHC}%
\label{sec:LHC}}

\begin{figure}[t]
\begin{center}
\includegraphics[width=6cm]{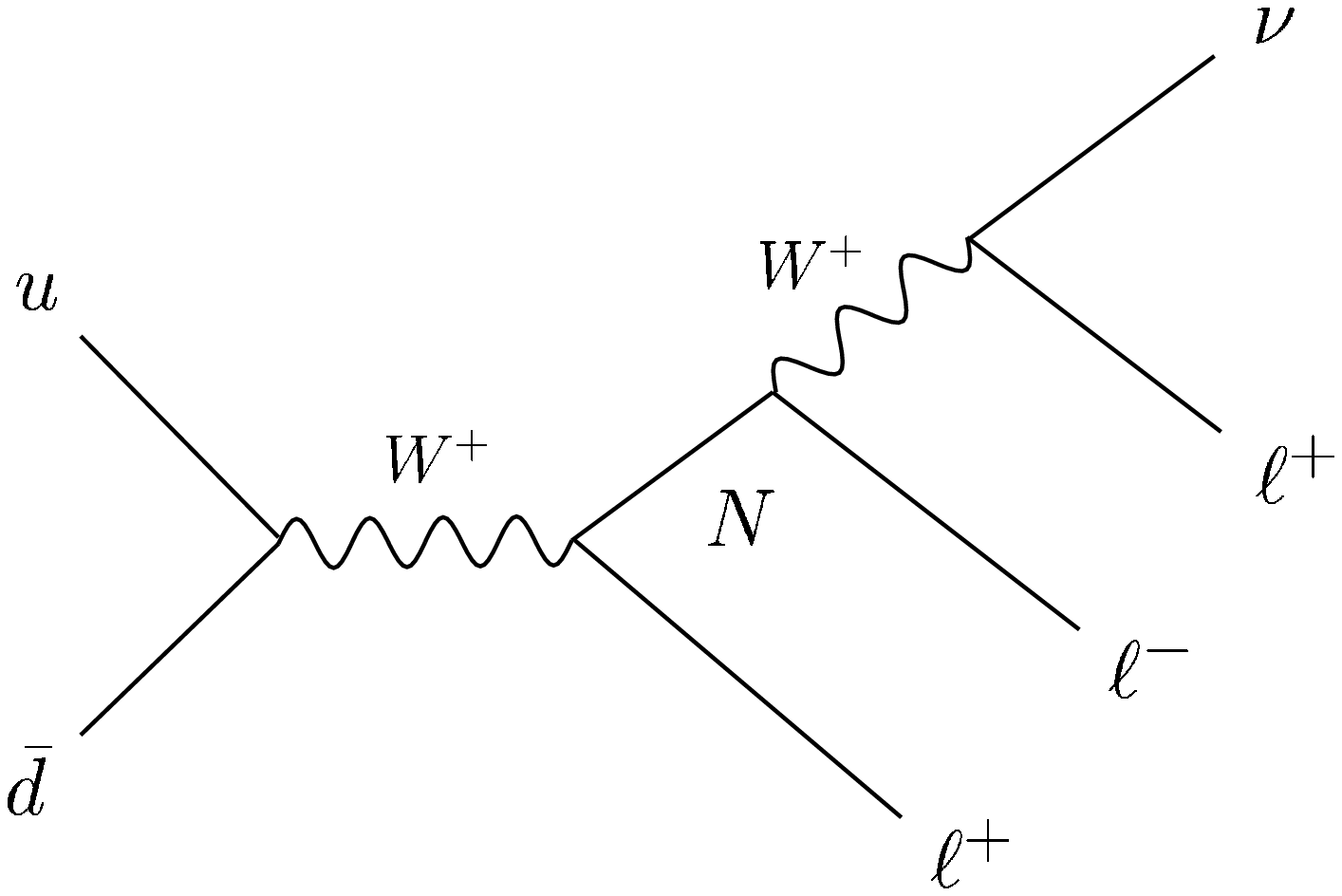}\hspace{15mm}
\includegraphics[width=6cm]{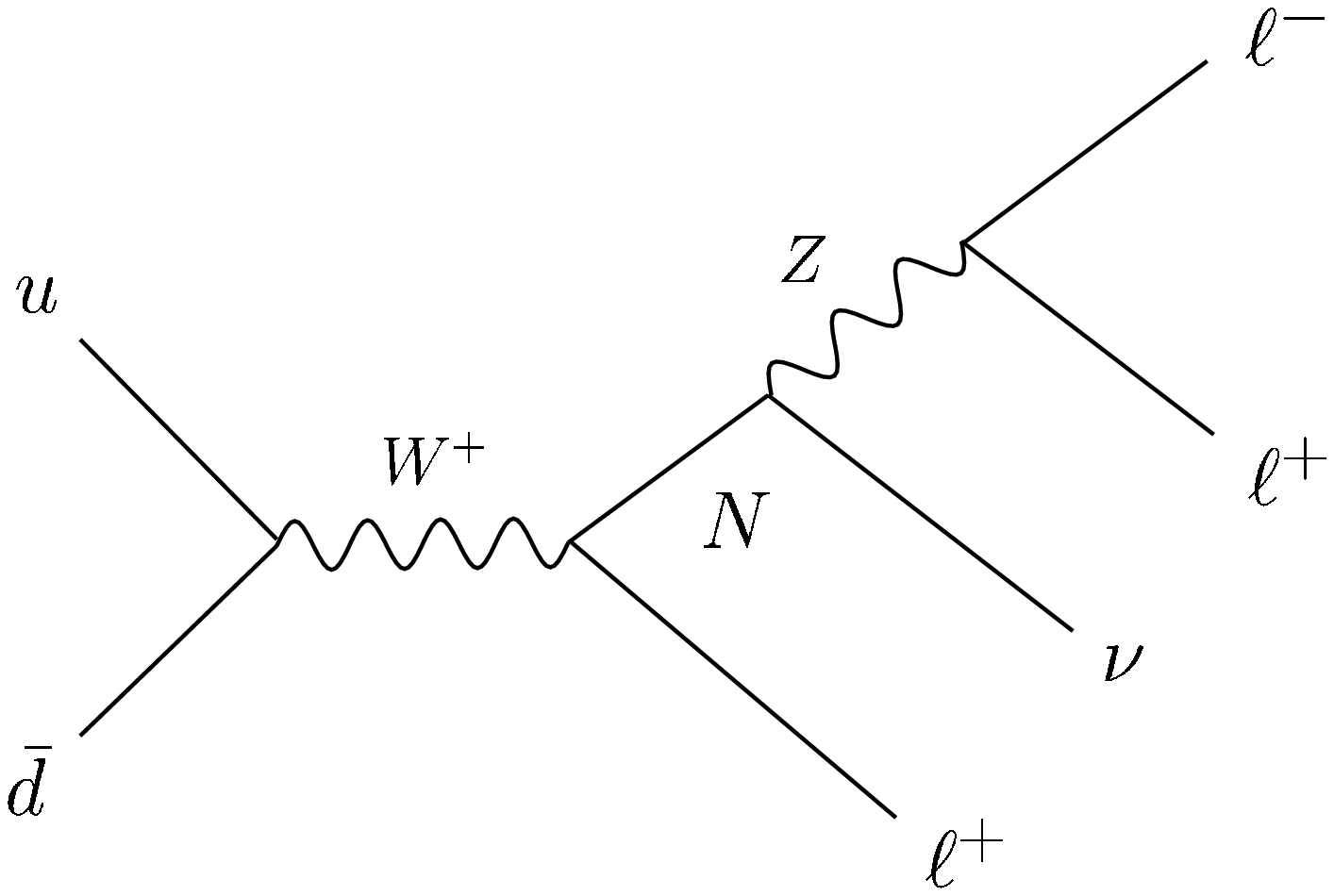}
\caption{The lepton number preserving tri-lepton processes at the 
LHC.\bigskip}
\label{fig:3leptons}
\end{center}
\end{figure}

The production of KK-excited neutrino states is the most important
signal in our scenarios, since the signal enables us to explore the
mechanism responsible for the generation of tiny neutrino masses. An
immediate question is which processes we should pay attention to find
out the signal at the LHC\@. As shown in the previous work~\cite{HMY},
the tri-lepton signal with missing transverse energy is most
prominent since only a small fraction of SM processes contributes to
the background against the signal. This lepton number conserving 
processes, $pp\to\ell^\pm\ell^\mp\ell^\pm\nu(\bar{\nu})$, dominantly
occur through the diagrams shown in FIG.~\ref{fig:3leptons}. In this
section, we investigate such seesaw signatures in Models~1 and 2,
which are presented in the previous section as typical examples of the 
observable seesaw. In the following simulation study, we assume, for
simplicity, that the bulk mass parameters of right-handed neutrinos
are common in flavor space and the complex phases vanish in the
orthogonal matrix $O$. With these assumptions, we perform the
numerical analysis for the tri-lepton signal in various mass
hierarchies of neutrino masses, that is, the normal, inverted, and
degenerated patterns.

\bigskip\bigskip\bigskip

%%%%%%%%%%%%%%%%%%%%%%%%%%%%%%%%%%%%%%%%%%%%%%%%%%%%%%%%%%%%%%%%%
{\centering\subsection{Constraints on Neutrino Yukawa Couplings}}

Before going to discuss the simulation study in details, we summarize
the neutrino mass and mixing matrices which are mandatory to
investigate the collider signatures at the LHC\@. The two matrices are
parameterized as
\begin{eqnarray}
 M_\nu^d &=&
 \left(\begin{array}{ccc}
   m_{\nu_1} & & \\
   & m_{\nu_2} & \\
   & & m_{\nu_3} 
 \end{array}\right),
 \qquad\quad
 \phi \;=\;
 \left(\begin{array}{ccc}
   e^{i\varphi_1} & & \\
   & e^{i\varphi_2} & \\
   & & 1
  \end{array}\right),
 \nonumber \\
 U_{\rm MNS} &=&
 \left(\begin{array}{ccc}
   c_{12}c_{13} &
   s_{12}c_{13} &
   s_{13}e^{-i\delta} \\
  -s_{12}c_{23} - c_{12}s_{23}s_{13}e^{i\delta} &
   c_{12}c_{23} - s_{12}s_{23}s_{13}e^{i\delta} &
   s_{23}c_{13} \\
   s_{12}s_{23} - c_{12}c_{23}s_{13}e^{i\delta} &
  -c_{12}s_{23} - s_{12}c_{23}s_{13}e^{i\delta} &
   c_{23}c_{13}
 \end{array}\right)\phi\,,
\end{eqnarray}
where $s_x$ $(c_x)$ means $\sin\theta_x$ ($\cos\theta_x$). The Dirac
and Majorana phases are denoted by $\delta$ and $\varphi_{1,2}$, 
respectively. Note that Majorana phases are not relevant in Model~2
since there is no Majorana mass term in the neutrino sector. The
neutrino mass differences and the generation mixing parameters have
been measured at neutrino oscillation
experiments~\cite{neu_analysis}. We take their typical values,
\begin{alignat}{2}
  \Delta m_{21} &\;\,\equiv\;\, m_{\nu_2}-m_{\nu_1} &\;\;=\; 
  9\times 10^{-3} ~\text{eV}, \\
  \Delta m_{32} &\;\,\equiv\; |m_{\nu_3}-m_{\nu_2}| &\;\;=\;
  5\times 10^{-2} ~\text{eV},
\end{alignat}
\begin{equation}
  s_{12} \,=\, 0.56, \qquad 
  s_{23} \,=\, 0.71, \qquad
  s_{13} \,\leq\, 0.22.
\end{equation}
The neutrino mass spectrum is allowed to have three different types of
hierarchies and is summarized in TABLE~\ref{table:hierarchy}, where we
define $m_{\rm tot}=(0.67\,\text{eV}-2\Delta m_{21}-\Delta m_{32})/3
\simeq0.2$~eV, taking account of the cosmological 
bound: $\sum_i m_{\nu_i}\leq0.67$~eV~\cite{mtotal}.

\begin{table}[t]
\begin{center}
\begin{tabular}{lcccc} \hline\hline 
  && $m_{\nu 1}$ & $m_{\nu 2}$ & $m_{\nu 3}$ \\ \hline
~Normal && 0 & $\Delta m_{21}$ & $\Delta m_{21} + \Delta m_{32}$ \\ 
~Inverted && ~ $\Delta m_{32}-\Delta m_{21}$ ~ & $\Delta m_{32}$ & 0 \\ 
~Degenerate ~ && $m_{\rm tot}$ & ~ $m_{\rm tot}+\Delta m_{21}$ 
& ~~ $m_{\rm tot}+\Delta m_{21}+\Delta m_{32}$ ~ \\ \hline\hline
\end{tabular}
\caption{\small Three types of neutrino mass hierarchies used in the
simulation study.\bigskip}
\label{table:hierarchy}
\end{center}
\end{table}   

Since the scenarios we are studying also affect several physical
observables such as the flavor-changing processes of charged
leptons~\cite{LFV}, it is important to consider the constraints on
neutrino Yukawa couplings to have proper representative
points. Integrating out all heavy KK neutrinos, we obtain the
following dimension 6 operator ${\cal O}^{(6)}$ in low-energy
effective theory which contributes to the leptonic flavor-changing
neutral current;
\begin{eqnarray}
  {\cal O}^{(6)} \,=\; \frac{1}{v^2}
  \big(\bar{L}\tilde H\big) \epsilon_N i \Slash{\partial}
  \big(\tilde H^\dagger L\big),
\end{eqnarray}
where the coefficient matrix $\epsilon_N$ ($\epsilon_N^{(1)}$ 
and $\epsilon_N^{(2)}$ for Models~1 and 2) turns out to be
\begin{eqnarray}
 \epsilon_N^{(1)} &=&
 \frac{2}{\delta_M} U_{\rm MNS}\,M_\nu^d\,U^\dagger_{\rm MNS}\,, \\[1mm]
 \epsilon_N^{(2)} &=& \frac{e^{-\pi Rm_d}}{2m_d^2}
 \Big[\cosh(\pi Rm_d)-\frac{\pi Rm_d}{\sinh(\pi Rm_d)}\Big]
 \,U_{\rm MNS} (M_\nu^d)^2 U^\dagger_{\rm MNS}\,.
\end{eqnarray}
The operator ${\cal O}^{(6)}$ receives phenomenological
constraints as shown in Ref.~\cite{lowene}, and each component of
neutrino Yukawa couplings is thus restricted by comparing the model
predictions of the coefficient with experimental data. In
particular, the most severe limit is given by the 1-2 component, i.e.,
the $\mu\to e\gamma$ search which puts on the upper bound more than 3
orders of magnitude stronger than the others. To weaken the bound on
this operator, especially for the 1-2 component, we take
representative values of lepton mixing matrix as shown in 
TABLE~\ref{table:MNS}\@. As a result, new physics parameters in the
coefficient $\epsilon_N$ such as $\delta_M$ in Model~1 and $Rm_d$ in
Model~2 turn out to be constrained as follows for each pattern of
neutrino mass hierarchy:
{\allowdisplaybreaks%
\begin{alignat}{3}
\text{(Model 1)} 
  & \qquad & 
  \delta_M &\;\geq\; 3.3~{\rm eV} & \quad & \text{for Normal}, \nonumber \\ 
  & \qquad & 
  \delta_M &\;\geq\; 4.4~{\rm eV} & \quad & \text{for Inverted}, \\
  & \qquad & 
  \delta_M &\;\geq\; 24.~{\rm eV} & \quad & \text{for Degenerate}, 
  \nonumber \\[3mm]
\text{(Model 2)} 
  & \qquad &
  -Rm_d &\;\leq\; 8.5-9.0 & \qquad & \text{for Normal}, \nonumber \\
  & \qquad &
  -Rm_d &\;\leq\; 8.5-8.9 & \qquad & \text{for Inverted}, \\
  & \qquad &
  -Rm_d &\;\leq\; 8.0-8.4 & \qquad & \text{for Degenerate}. \nonumber
\end{alignat}}%
Notice that the coefficient $\epsilon_N$ in Model~1 is irrelevant to
the compactification radius $R$ and so the bounds 
on $\delta_M$ are. For Model~2, the above bounds are obtained 
for $1/R=100-350$~GeV\@. The compactification radius is also limited
by the LEP experiment through the masses of KK excited neutrinos. The
lightest ones are $M_1=1/(2R)$ for Model~1 
and $M_1=\sqrt{1/R^2+m_d^2}$ for Model~2, and these states have not
be experimentally detected so far. We numerically checked that the
constraint is not so severe if $M_1>150$~GeV\@. Finally, the SM Higgs
mass is to be $m_h=120$ GeV in evaluating the decay widths of heavy
KK neutrinos. 

\begin{table}[t]
\begin{center}
\begin{tabular}{lcccccc} \hline\hline
& \multicolumn{3}{c}{Model 1} && \multicolumn{2}{c}{Model 2} \\ \hline 
& $s_{13}$ & $\,\delta\,$ & ~ $\varphi_{1,2}$ ~ && $s_{13}$ 
& $\delta$ ~ \\ \hline
~Normal   & 0.07 & $\pi$ & 0 && ~ 0.006 ~ & $\pi$ ~ \\
~Inverted & ~ 0.09 ~ & 0 & 0 && 0.19 & 0 ~ \\
~Degenerate ~ & 0.04 & $\pi$ & 0 && 0.05 & $\pi$ ~ \\ \hline\hline
\end{tabular}
\caption{\small The representative points for $U_{\rm MNS}$ in 
Models~1 and 2\@. The Majorana phases in $U_{\rm MNS}$ have no physical
relevance in the present work and are set to be zero.\bigskip}
\label{table:MNS}
\end{center}
\end{table}   

\bigskip\bigskip\bigskip

%%%%%%%%%%%%%%%%%%%%%%%%%%%%%%%%%%%%%%%%%%%%%%%%%%%%%%%%%%%%%%%%%
{\centering\subsection{Tri-lepton Signals at the LHC}}

\begin{figure}[t]
\begin{center}
\includegraphics[width=15.5cm]{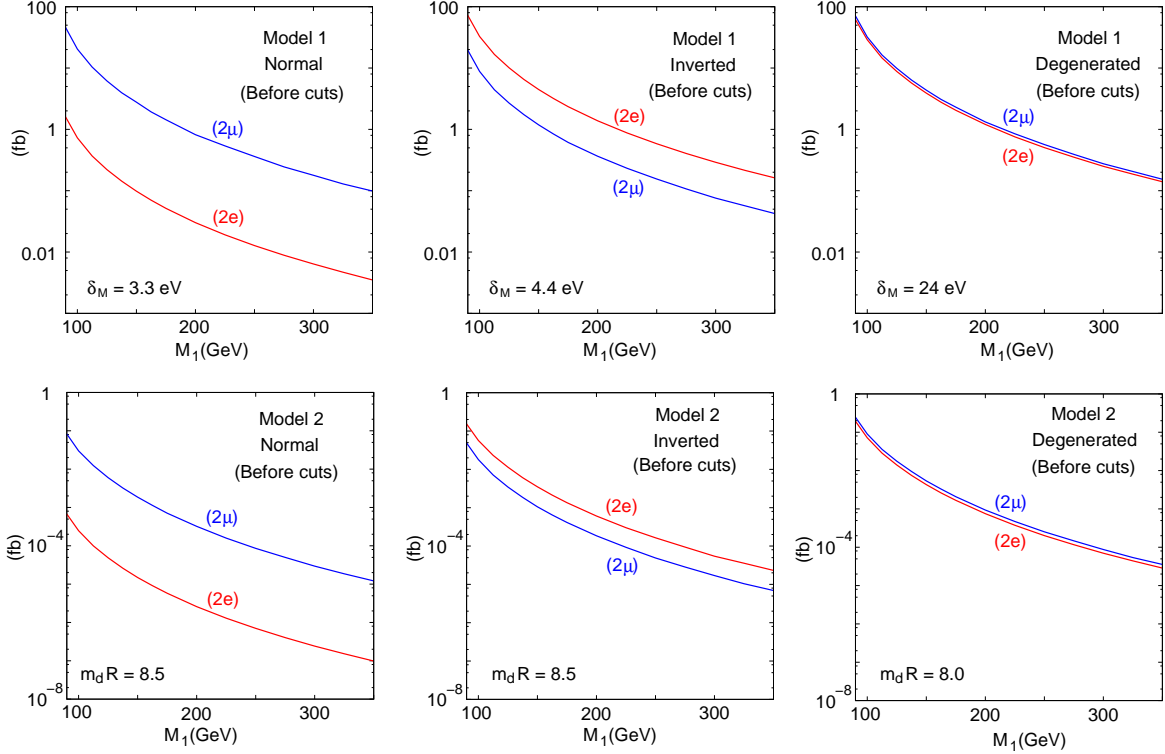}
\caption{Total cross sections of tri-lepton signals with fixed values
of $\epsilon_N$. The horizontal axes are the masses of the lightest
KK-excited modes. The upper and lower panels are for Models~1 and 2,
respectively.\bigskip}
\label{fig:Xsec_bc}
\end{center}
\end{figure}

Now let us investigate the tri-lepton signal of heavy neutrino
productions at the LHC\@. Since the tau lepton is hardly detected
compared to the others, we consider the signal event including only
electrons and muons. There are four kinds of tri-lepton 
signals: $eee$, $ee\mu$, $e\mu\mu$, and $\mu\mu\mu$. In this work, we
use two combined signals which are composed 
of $eee+ee\mu$ (the $2e$ signal) 
and $e\mu\mu+\mu\mu\mu$ (the $2\mu$ signal). Figure~\ref{fig:Xsec_bc}
shows the total cross sections for these signals from the 1st KK
neutrino production at the LHC, which are described as the functions
of their mass eigenvalues with fixed values of $\epsilon_N$. It is
found from the figure that the cross sections have the universal
behaviour within extra-dimensional models; for the normal mass 
hierarchy, the cross section for the $2\mu$ signal is about one or two
orders of magnitude larger than the $2e$ signal, and for the inverted
and degenerate spectra, the $2e$ signal cross section becomes larger
than or almost equal to the $2\mu$ one. Further, the cross section for
Model~2 is found to be small compared with that for Model~1. This is
due to a small wavefunction factor of low-lying KK neutrino 
mode, $f_1\sim 1/(Rm_d)$, which is suppressed by $\ln(M_\nu/v)$ to
have tiny neutrino masses. We have also evaluated the contributions of
tri-lepton signals from heavier KK neutrinos and found that they are
small by more than one order of magnitude and are out of reach of the
LHC experiment. A high luminosity collider with clean environment such
as the International Linear Collider would distinctly discover the
signatures of KK mode resonances.

\begin{figure}[t]
\begin{center}
\includegraphics[width=15.5cm]{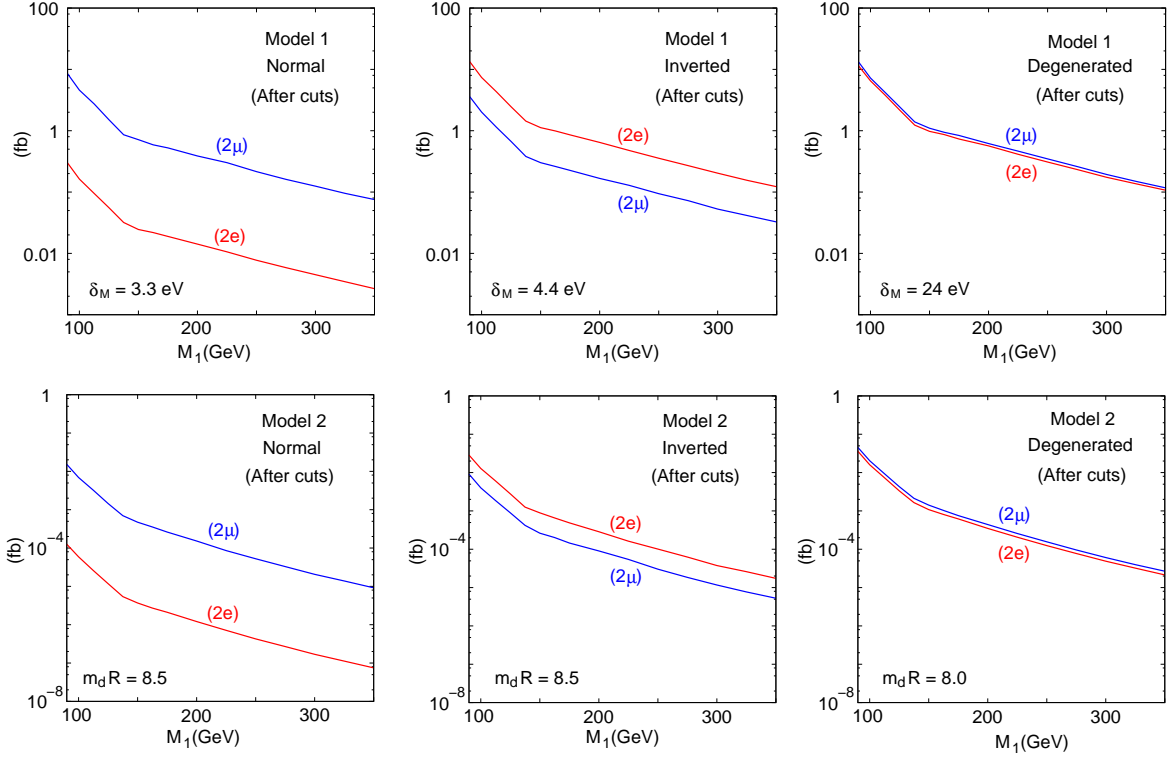}
\caption{Total cross sections of tri-lepton signals with fixed values
of $\epsilon_N$ after implementing the kinematical cuts (see the
text). The horizontal axes are the masses of the lightest KK-excited
modes. The upper and lower panels are for Models~1 and 2,
respectively.\bigskip}
\label{fig:Xsec_ac}
\end{center}
\end{figure}

To clarify whether the tri-lepton signal is captured at the LHC, it is
important to estimate SM backgrounds against the signal. The SM
backgrounds which produce or mimic the tri-lepton final state have
been studied~\cite{cut,cut2} and for the present purpose a useful
kinematical cut is discussed to reduce these SM
processes~\cite{cut2}. According to that work, we adopt the following
kinematical cuts for both Models:
(i) the existence of two like-sign charged 
leptons $\ell_1^\pm$, $\ell_2^\pm$, and an additional one with the
opposite charge $\ell_3^\mp$, 
(ii) both of the energies of like-sign leptons are larger than 30 GeV,
and 
(iii) the invariant masses from $\ell_1$ and $\ell_3$ and 
from $\ell_2$ and $\ell_3$ are larger than $m_Z+10$ GeV or smaller
than $m_Z-10$ GeV\@. The last condition is imposed to reduce the large
background from the leptonic decays of $Z$ bosons in the SM 
processes. Figure~\ref{fig:Xsec_ac} shows the total cross sections of
signals after imposing these kinematical cuts. To estimate the
efficiency for the signal events due to the cuts, we use the Monte
Carlo simulation using the CalcHep code~\cite{CalcHep}. Since the
event numbers of SM backgrounds after the cuts are about 260 for 
the $2e$ signal and 110 for the $2\mu$ one with the luminosity of
30~fb$^{-1}\,$\cite{cut2}, the $2\mu$ events are expected to be
observed if the lightest KK mass $M_1$ is less than a few hundred
GeV\@.

\begin{figure}[t]
\begin{center}
\begin{minipage}{1.0\linewidth}
\includegraphics[width=15.5cm]{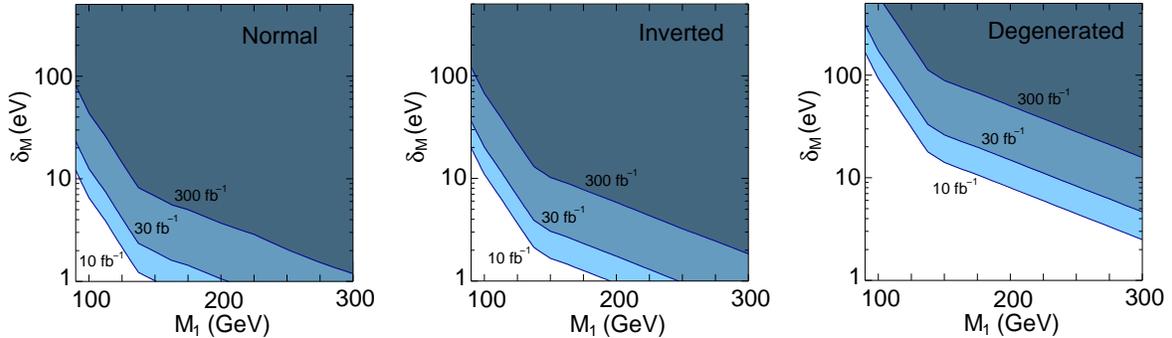}
\caption{Luminosity for the 3$\sigma$ reach in Model~1. The 10, 30,
and 300 fb$^{-1}$ contours are shown on the $(M_1,\delta_M)$ plane for
three types of neutrino mass hierarchy.\bigskip}  
\label{fig:Reach1}
\end{minipage}
\end{center}
\end{figure}

For Model~1, FIG.~\ref{fig:Reach1} shows the luminosity which is
required to find the seesaw neutrino signal at the LHC as the
contour plots on the parameter plane. The luminosity contours for 10,
30, and 300 fb$^{-1}$ are depicted in the figures. These contours are
obtained by computing the significance for the signal discovery
\begin{eqnarray}
  \qquad
  S \,=\, \sqrt{S_e^2 + S_\mu^2}\,, \qquad\qquad
  S_i \,=\, \frac{N_{S_i}}{\sqrt{N_{S_i}+N_{B_i}}} \qquad (i=e,\mu),
\end{eqnarray}
where $N_S$ ($N_B$) denotes the total number of 
the 2$e$ or 2$\mu$ events (that of the corresponding SM backgrounds)
after the kinematical cuts. Since both $N_S$ and $N_B$ are
proportional to the luminosity, it is possible to estimate the
required luminosity for, e.g.\ giving $S=3$, which is plotted in the
above figures. The luminosity for signal confirmation ($S\geq5$) are
also found by rescaling the results, according to the formula
(luminosity)~$\propto S^2$. It is found that, if $M_1$ is less than a
few hundreds GeV, the signals would be observed at an early run of the
LHC, in particular, Model~1 with the degenerate mass spectrum will
definitely be excluded or confirmed. A larger luminosity is needed for
a smaller size of extra dimension to reveal its existence.

Model~2, as it stands, generally predicts too small production rates
to be found out at the LHC\@. However the extra-dimensional framework
has various options without introducing additional particles. That
leads to simple modifications of the model and can make it
observable, as explained explicitly in the next section.

\bigskip\bigskip\bigskip

%%%%%%%%%%%%%%%%%%%%%%%%%%%%%%%%%%%%%%%%%%%%%%%%%%%%%%%%%%%%%%%%%
{\centering\section{Other Configurations for Observable Seesaw}%
\label{sec:others}}

We have discussed the collider signatures of two typical models of
observable seesaw. They are constructed in five-dimensional spacetime
and utilize the mechanisms which are peculiar to the presence of extra
dimensions; the specific value of bulk Majorana mass in Model~1 and
the suppression factor from localized wavefunction in Model~2. As we
mentioned before, while the Lagrangian is simple and common, the
five-dimensional theory makes right-handed neutrinos observable in
various ways which have their own physical meanings. Among
them, we here present three possibilities; localized Majorana mass
terms, boundary conditions of bulk fields, and curved gravitational
backgrounds (and their combinations).

\bigskip\bigskip\bigskip

%%%%%%%%%%%%%%%%%%%%%%%%%%%%%%%%%%%%%%%%%%%%%%%%%%%%%%%%%%%%%%%%%
{\centering\subsection{Boundary Majorana Masses}%
\label{sec:b_mass}}

We first consider the case that Majorana mass parameters for bulk
fermions depend on the extra-dimensional coordinate $x^5$. An
interesting case is that Majorana masses are localized at the
boundaries of fifth dimension ($x^5=0,\,\pi R$). The boundary Majorana
masses may be natural if the lepton number symmetry is exact in the
bulk and locally broken at the boundaries. For the flat background,
the two fixed points are physically equivalent, and in the following
we choose $x^5=0$ as an example. We consider the Lagrangian
\begin{eqnarray}
  {\cal L} \;=\; i\overline\Psi\partial\hspace*{-2.2mm}/\Psi
  -\overline\Psi m_d\theta(x^5)\Psi
  -\Big[\,\frac{1}{2}\overline{\Psi^c}M\Psi
  +\overline{\Psi} mL +\text{h.c.}\Big]\delta(x^5).
\end{eqnarray}
If one also includes the $M_5$ term, $M$ is replaced with $M+M_5$ in
the following formulas. The solutions to the bulk equations of motion
have been given in Section~\ref{sec:Model2}\@. The mass matrices in
four-dimensional effective theory are found
\begin{eqnarray}
  M_{K_{mn}} \,=\; -\frac{n}{R}\delta_{mn}, \quad\;
  M_{R_{mn}} \,=\; \frac{f_mf_n}{\pi R}M, \quad\;\;
  M_{L_{mn}} \,=\; 0, \quad\;
  m_n \;=\; \frac{f_n}{\sqrt{\pi R}}\,m.
\end{eqnarray}
One type of the Majorana masses, $M_L$, vanishes since $N_L^{\,n}$
have the negative $Z_2$ parity and the wavefunctions become zero at the
boundary. Another Majorana mass matrix, $M_R$, has the off-diagonal
($m\neq n$) entries since the KK momentum is not conserved at the
boundary.

It seems difficult to diagonalize the heavy-field mass 
matrix $M_N$ which is composed of $M_{R,L}$ and the KK 
masses $M_K$. However the mixing vertex $V$ between heavy modes and
the SM sector can be evaluated by getting the inverse of $M_N$ that is
given by
\begin{eqnarray}
  V \;=\; U_\nu^\dagger \big(%
  \begin{array}{ccccc}
  \!\tfrac{\sqrt{\pi R}}{-f_0}\,m^\dagger M^{-1}
  & \,0 & \,0 & \,0 & \,\cdots
  \end{array}\big)U_N.
\end{eqnarray}
Notice that all the components but the 1st one are vanishing in the
interaction basis. From this mixing matrix, we obtain the
seesaw-induced neutrino masses and the heavy-mode mixing with the SM
neutrinos:
\begin{eqnarray}
  M_\nu &=& m^{\rm t}M^{-1\,*}m,
  \label{Mnu_other1}  \\[1mm]
  \nu \; &=& U_\nu\nu_d -\frac{\sqrt{\pi R}}{f_0}\,m^\dagger
  M^{-1}\epsilon N_R^{\,0\,*}.
  \label{nu_other1}
\end{eqnarray}
The form of $M_\nu$ is apparently the same as the usual
four-dimensional seesaw mechanism, but $\nu$ has an extra 
factor $\sqrt{\pi R}/f_0$ which originates from the extra dimension,
i.e.,\ the volume factor and the localization factor controlled by the
bulk Dirac mass $m_d$. These two factors are available in the case of
boundary Majorana masses and are not obtained for bulk Majorana masses
as in the previous section. This is because these factors appear twice
in the boundary Majorana masses for bulk fields, but only once in the
boundary coupling to the SM fields, which result in the cancellation
only for the seesaw-induced masses. In the present case, the two
factors can be used for enhancing the heavy-mode couplings to the SM
neutrinos, while keeping the seesaw-induced masses un-affected and
made tiny. The enhancement by the localization factor 
requires $-Rm_d\gg1$. If one introduced Majorana masses at another
boundary $x^5=\pi R$, the bulk mass parameter $m_d$ should be replaced
with $-m_d$ in the formula.

In this way, the Majorana masses on the boundary realize an observable
seesaw model with appropriate wavefunction factors. The SM fields
(neutrinos, electroweak gauge bosons, etc.) interact with the bulk
sector only through the zero mode $N_R^{\,0}$. It is noticed 
that $N_R^{\,0}$ is not a mass eigenstate, and the cross
sections for collider physics might be peaked at the mass eigenvalues
of KK neutrinos via the mixing with $N_R^{\,0}$. This is however not
the case in a quantitative meaning: the KK-mode contamination 
of 1\% mixing is found from \eqref{nu_other1} to 
imply $\sqrt{R}m/(f_0M)\sim{\cal O}(10^{-2})$. This in turn implies
by the seesaw formula \eqref{Mnu_other1} that the Majorana mass
parameter for $N_R^{\,0}$ is roughly given by $10^4\times M_\nu$ and
very small. Therefore the heavy KK neutrinos do not so much mix with
such a light zero mode and cannot be detected at collider experiments.

The conclusion is that, in the extra-dimensional setup in this
subsection, the zero-mode wavefunction factors enhance the heavy-mode
couplings, keeping the usual seesaw formula, and play a key role for
realizing the observable seesaw. However from a phenomenological
viewpoint, only the light zero mode is found to be accessible. That
depends on which elements are vanishing in the neutrino mass matrices
and could be changed by some effects within the model or its
extensions. For example, an additional boundary mass or interaction
term would lead to a repulsive effect which makes $Z_2$-odd fields off
from the boundary so that they obtain nonzero Majorana masses. Another
option is that a singular boundary profile is regulated by introducing
some scalar field, which generates a four-dimensional domain wall with
a finite width along the extra dimension. That would lead to
non-vanishing couplings of bulk fields in four-dimensional effective
theory.

\bigskip\bigskip\bigskip

%%%%%%%%%%%%%%%%%%%%%%%%%%%%%%%%%%%%%%%%%%%%%%%%%%%%%%%%%%%%%%%%%
{\centering\subsection{Boundary Conditions}%
\label{sec:b_cond}}

Another important option of five-dimensional theory is to choose
boundary conditions for bulk fields. For a finite size of extra space,
the boundary conditions determine the bulk profile, i.e.\
wavefunctions of higher-dimensional fields, and then fix their
low-energy physics. We have so far discussed the standard boundary
condition for a five-dimensional spinor $\Psi$ on 
the $S^1/Z_2$ orbifold, that is, the Neumann and Dirichlet type
boundary conditions for the upper and lower components, 
respectively; $\partial_5\chi_R^n=\chi_L^n=0$ at 
both $x^5=0$ and $\pi R$\@. In this section, let us consider another
mixed-type condition:\footnote{An overall sign is fixed by assuming
that the upper component of $\Psi$ has non-vanishing wavefunction
at the boundary $x^5=0$ where the SM fields reside.}
\begin{eqnarray}
  \Psi(-x^5) &=& +\gamma_5\Psi(x^5), \nonumber \\
  \Psi(-x^5+2\pi R) &=& -\gamma_5\Psi(x^5),
  \label{b_cond}
\end{eqnarray}
i.e.\ the upper component has a positive (negative) parity under the
reflection about the $x^5=0$ ($x^5=\pi R$) boundary. The lower
component has the opposite parity assignment. In terms of KK-mode 
wavefunctions, $\partial_5\chi_R^n(0)=\chi_L^n(0)=0$ and 
$\chi_R^n(\pi R)=\partial_5\chi_L^n(\pi R)=0$, in the absence of
extra boundary terms. Notice that this is equivalent to the
Scherk-Schwarz boundary condition~\cite{SS} where a non-trivial twist
is imposed in circulating along the extra 
dimension: $\Psi(x^5+2\pi R)=-\Psi(x^5)$.

Let us consider the following Lagrangian
\begin{eqnarray}
  {\cal L} \;=\;
  i\overline\Psi\partial\hspace*{-2.2mm}/\Psi 
  -\Big[\,\frac{1}{2}\overline{\Psi^c}M\Psi
  +\overline{\Psi} mL\delta(x_5)+\text{h.c.}\Big],
\end{eqnarray}
and evaluate the seesaw mass matrix under the boundary
conditions~\eqref{b_cond}. The wavefunctions for free bulk fields are
given by
\begin{eqnarray}
  \quad
  \chi_R^n \;=\; \frac{1}{\sqrt{\pi R}}
  \cos\bigg[\frac{(n-\frac{1}{2})}{R}x^5\bigg], \qquad
  \chi_L^n \;=\; \frac{1}{\sqrt{\pi R}}
  \sin\bigg[\frac{(n-\frac{1}{2})}{R}x^5\bigg]. \qquad (n\geq1)
  \label{wave_other2}
\end{eqnarray}
The mass matrices in four-dimensional effective theory are found
\begin{eqnarray}
  M_{K_{mn}} \,=\; -\,\frac{n-\frac{1}{2}}{R}\delta_{mn}, \quad\;
  M_{R_{mn}} \,=\; M_{L_{mn}} \,=\; M\delta_{mn}, \quad\;\;
  m_n \;=\; \frac{m}{\sqrt{\pi R}}\,.
\end{eqnarray}
The only difference from the previous standard boundary condition is
the KK mass spectrum $M_K$. We find the seesaw-induced neutrino mass
and the heavy-mode mixing with the SM neutrinos:
\begin{eqnarray}
  M_\nu &=& \frac{1}{2\pi R}\,m^{\rm t}
  \frac{\pi R|M|}{\coth(\pi R|M|)}\,\frac{1}{M^*}\,m, 
  \label{Mnu_other2}  \\[1mm]
  \nu &=& U_\nu\nu_d \,-\,\frac{m^\dagger}{\sqrt{\pi R}}\,
  \sum_{n=1}\frac{1}{|M|^2\!+\big(\frac{n-\frac{1}{2}}{R}\big)^2}
  \,\bigg[\frac{n-\frac{1}{2}}{R}\,N_L^{\,n}+M^*
  \epsilon N_R^{\,n\,*}\bigg].
  \label{nu_other2}
\end{eqnarray}
The light neutrino mass $M_\nu$ has the 
factor $\pi R|M|/\coth(\pi R|M|)$ as a consequence of summing up
the heavy-mode seesaw contributions. Notice that, for the standard
boundary condition, this factor is $\pi R|M|/\tanh(\pi R|M|)$. The
difference is understood in the following two limits: For the large
radius limit, $RM\gg1$, the two boundaries are so separated in the
extra-dimensional space that the difference of boundary conditions 
at $x^5=\pi R$ is irrelevant to the SM physics at $x^5=0$, and two
factors merge into the same value $\pi R|M|$. The other 
case, $RM\ll1$, is the decoupling limit of KK modes. They become so
heavy that the low-energy physics is determined by light modes
only. It is the chiral zero mode in case of the standard boundary
condition. For the present twisted boundary condition, the zero mode
is absent and the limit $RM\to0$ leads to vanishing seesaw-induced
masses. That is, the inverse seesaw suppression~\cite{inverse_seesaw}
is realized at each KK level and the total seesaw-induced mass is
proportional (not inverse proportional) to heavy-field Majorana 
mass $M$.

In this way, the boundary condition mechanism leads to the situation
that no massless mode appears in the KK decomposition and therefore
bulk Majorana masses can be made small without being conflicting with
the heavy-mode integration. Let us consider the case of small Majorana
masses ($RM\ll1$). The seesaw-induced mass and the mass eigenvalues
of KK Dirac neutrinos become 
\begin{eqnarray}
  M_\nu \;\simeq\; \frac{\pi R}{2}m^{\rm t}Mm, \qquad\quad
  M_n \;\simeq\; \frac{n-\frac{1}{2}}{R} \qquad (n\geq1).
\end{eqnarray}
This agrees with the spectrum of Model~1 discussed in 
Section~\ref{sec:Model1} with the 
replacement $\delta_M\leftrightarrow M$. The mixing with heavy modes
also has the correspondence under this replacement and with a field
rearrangement. Therefore the present model with the twisted boundary
condition is observable and gives the same seesaw phenomenology, in
particular the LHC signatures, as given in Section~\ref{sec:LHC}\@. A
difference of two models is the interpretation of small 
parameters $\delta_M$ and $M$. The parameter $\delta_M$ in Model~1
is a tiny deviation from the fixed value of model 
parameter ($M_5=\frac{1}{2R}$) and is hard to be determined in
dynamical way. On the other hand, $M$ is a Lagrangian parameter itself
and is easier to be suppressed and controlled with high-energy physics.

If one includes the bulk Dirac mass $m_d$, the above formulas in
low-energy effective theory are modified as
\begin{eqnarray}
  M_\nu &=& \frac{\pi R}{2}\,m^{\rm t} \bigg[ -\pi Rm_d 
  +\frac{\pi R\sqrt{|M|^2+m_d^2\,}}{\tanh\!\big(\pi R\sqrt{|M|^2+
   m_d^2\,}\big)}\bigg]^{-1}\!\!M\,m,  \\[1mm]
   \nu &\simeq& U_\nu\nu_d \,-\,\frac{m^\dagger}{\sqrt{\pi R}}\,
  \sum_{n=1}\frac{(n-\frac{1}{2})/R}{\big[\big(n-
   \frac{1}{2}\big)/R\big]^2\!+m_d^2}\,N_L^{\,n}.
\end{eqnarray}
In the regime $-Rm_d\gg1$, the Dirac mass parameter is effective in
suppressing the seesaw-induced masses $M_\nu$, compared 
with \eqref{Mnu_other2}: for small bulk Majorana masses, we 
obtain $M_\nu\simeq m^{\rm t}(M/4m_d)m$.

\bigskip\bigskip\bigskip

%%%%%%%%%%%%%%%%%%%%%%%%%%%%%%%%%%%%%%%%%%%%%%%%%%%%%%%%%%%%%%%%%
{\centering\subsection{Boundary Majorana Masses and Boundary Conditions}}

An interesting and physically different scheme is given by considering
both of boundary Majorana mass and non-trivial boundary condition of
bulk neutrinos, discussed in the previous two sections. This model is
particular in that the seesaw-induced neutrino mass vanishes for any
values of model parameters. Therefore the heavy-mode couplings to the
SM sector are arbitrarily fixed so that the scenario is observable at
collider experiments. The Majorana mass parameters do not appear in
any place of low-energy effective theory at the leading order.

Let us consider the same Lagrangian as in Section~\ref{sec:b_mass}
\begin{eqnarray}
  {\cal L} \;=\; i\overline\Psi\partial\hspace*{-2.2mm}/\Psi
  -\Big[\,\frac{1}{2}\overline{\Psi^c}M\Psi
  +\overline{\Psi} mL
  +\text{h.c.}\Big]\delta(x^5).
\end{eqnarray}
That is, the Majorana masses for bulk fermions are only on the SM
boundary. Further we assume the twisted boundary condition as in
Section~\ref{sec:b_cond}:
\begin{eqnarray}
  \Psi(-x^5) &=& +\gamma_5\Psi(x^5), \nonumber \\
  \Psi(-x^5+2\pi R) &=& -\gamma_5\Psi(x^5).
\end{eqnarray}
Therefore the wavefunctions and KK masses are given 
by~\eqref{wave_other2} as previously. The mass matrices in
four-dimensional effective theory are found
\begin{eqnarray}
  M_{K_{mn}} \,=\; -\,\frac{n-\frac{1}{2}}{R}\,\delta_{mn}, \quad\;
  M_{R_{mn}} \,=\; \frac{1}{\pi R}\,M, \quad\;
  M_{L_{mn}} \,=\; 0, \quad\;
  m_n \,=\, \frac{m}{\sqrt{\pi R}}\,.
\end{eqnarray}
The Majorana masses $M_L$ vanish since $N_L^{\,n}$ have the
negative $Z_2$ parity and the wavefunctions become zero at 
the $x^5=0$ boundary on which the Lagrangian mass term is
placed. Another Majorana mass matrix $M_R$ takes the common value for
all the matrix elements including the off-diagonal ones. The vertex
matrix $V$ of heavy modes can be evaluated by taking the inverse 
of $M_N$ that is given by 
\begin{eqnarray}
  V \;=\; -\sqrt{\tfrac{4R}{\pi}}\,U_\nu^\dagger m^T\big(%
  \begin{array}{cccccccc}
  0\, & 1\, & 0\, & \frac{1}{3}\, & 0\, & \frac{1}{5}\, & 0\, & \cdots
  \end{array}\big)U_N.
\end{eqnarray}
Notice that the ($2n-1$)-th components are all vanishing in the
interaction basis of KK modes. Further the non-vanishing elements do
not depend on the Majorana mass parameter $M$. From this mixing
matrix, we find the seesaw-induced neutrino mass and the heavy-mode
mixing with the SM neutrinos:
\begin{eqnarray}
  M_\nu &=& 0, \\[1mm]
  \nu \; &=& U_\nu\nu_d -\sqrt{\frac{R}{\pi}}\,m^\dagger
  \sum_{n=1}\frac{2}{2n-1}\,N_L^{\,n}.
\end{eqnarray}
It is interesting that the light neutrino mass $M_\nu$ vanishes,
irrespectively of model parameters. The heavy-mode mixing is governed
by the compactification scale and the boundary mass $m$. Their ratio
can therefore be arbitrarily fixed and made sizable. In this model,
the bulk Majorana mass $M$ does not join in any formula of the seesaw
operation and only affects the mass spectrum of heavy modes. The
spectrum is found to be roughly determined only by the
compactification scale and may be corrected by Majorana masses which
are suppressed by the cutoff scale of the theory.

The above result shows that the scheme in this subsection gives a
natural realization of the observable seesaw in the zero-th
approximation. Towards a phenomenologically viable model, nonzero
neutrino masses are needed to be generated by some dynamics. Among
various possibilities, a simple way is to put, as a correction, the
Majorana masses in the bulk and/or on the other boundary $x^5=\pi R\,$:
\begin{eqnarray}
  \Delta{\cal L} \;=\; -\frac{1}{2}\big(\overline{\Psi^c}
  M_b\Psi+\text{h.c.}\big) -\frac{1}{2}\big(\overline{\Psi^c}
  M_\pi\Psi+\text{h.c.}\big)\delta(x^5-\pi R).
\end{eqnarray}
Repeating the previous procedure with these terms, we obtain the
seesaw-induced neutrino masses
\begin{eqnarray}
  \Delta M_\nu \;=\; \frac{1}{2\pi R}\,m^{\rm t}\,
  (\pi R)^2\Big(M_b+\frac{1}{\pi R}M_\pi\Big)m
  \,+{\cal O}(RM_b,\,RM_\pi).
\end{eqnarray}

Finally we briefly comment on other patterns of the model. There
seems to exist 3 degrees of freedom: the boundary Majorana masses 
on $x^5=0$ or $\pi R$, the SM fields reside at $x^5=0$ or $\pi R$, and
the choice of boundary condition (the overall sign of $Z_2$ parity
assignment). However an actual freedom is only one, because two
boundaries are equivalent in the flat background, and the upper and
lower components of bulk fermions can be appropriately exchanged. As
the remaining freedom, let us consider the situation that boundary
Majorana masses are placed at $x^5=\pi R$, instead of $x^5=0$
discussed before. The boundary condition is the same as previously
and we then find 
\begin{eqnarray}
  M_\nu &=& \frac{1}{2\pi R}\,m^{\rm t}
  \Big(\frac{\pi RM}{2}\Big)\,m,  \\[1mm]
  \nu \; &=& U_\nu\nu_d -\sqrt{\frac{R}{\pi}}\,m^\dagger\,
  \sum_{n=1}\bigg[\,\frac{2}{2n-1}\,N_L^{\,n}
  -\frac{(-1)^n}{2n-1}M^*\,\epsilon
  N_R^{\,n\,*}\bigg].
\end{eqnarray}
The contribution of $N_L^{\,n}$ does not depend on the Majorana mass
parameter $M$, and therefore the observable seesaw is realized for a
suitably value of $M$ for obtaining tiny seesaw-induced masses, while
keeping the $N_L^{\,n}$ mixing sizable. The heavy neutrinos are
degenerate in mass and their spectrum is almost given by the KK 
masses $(n-\tfrac{1}{2})/R\,$ ($n\geq1$).

\bigskip\bigskip\bigskip

%%%%%%%%%%%%%%%%%%%%%%%%%%%%%%%%%%%%%%%%%%%%%%%%%%%%%%%%%%%%%%%%%
{\centering\subsection{AdS$_5$ Gravitational Background}}

So far we have discussed the bulk Lagrangian on the flat gravitational
background. Another typical geometry of extra dimension is given by
the so-called AdS$_5$ warped background~\cite{RS}. That is a solution
of the Einstein equation in the five-dimensional theory with
appropriately tuned cosmological constants both in the bulk and on the
boundaries. The line element is
\begin{eqnarray}
  ds^2 \;=\; e^{-2k|x^5|}\eta_{\mu\nu}dx^\mu dx^\nu-dx_5^2,
\end{eqnarray}
where $k$ is the AdS curvature and is related to the bulk cosmological
constant. Neutrino physics on the warped geometry has been studied
in the absence of Majorana masses~\cite{RS_neutrino}. Let us consider
the bulk field Lagrangian on this background. Evaluating the spin
connection and normalizing kinetic 
term ($\Psi\to e^{\frac{3}{2}k|x^5|}\Psi$), we have
\begin{eqnarray}
  {\cal L} \;=\; i\bar\Psi\Slash\partial\Psi
  -e^{-k|x^5|}\Big[\bar\Psi\gamma_5\partial_5\Psi
  +\overline\Psi \Big(m_d-\frac{k}{2}\gamma_5\Big)\theta(x^5)\Psi
  +\frac{1}{2}\big(\overline{\Psi^c}
  M\Psi+\text{h.c.}\big)\Big]. \quad
\end{eqnarray}
The bulk Dirac and Majorana mass terms depend on the extra-dimensional
coordinate $x^5$ which arise from the warped metric. In the absence
of bulk mass terms, the solutions to the bulk equations of motion are
given by
\begin{eqnarray}
  \chi^n_R &=& \frac{h_n}{\sqrt{\pi R}}\,e^{\frac{1}{2}k|x^5|}
  \cos\!\Big[\frac{nh_n^2}{kR}
  \big(e^{k|x^5|}-1\big)\Big],  \\
  \chi^n_L &=& \frac{h_n}{\sqrt{\pi R}}\,e^{\frac{1}{2}k|x^5|}
  \sin\!\Big[\frac{nh_n^2}{kR}\big(e^{k|x^5|}-1\big)\Big],
\end{eqnarray}
where the normalization factor is given by $h_n=
\sqrt{\pi kR/2^{\delta_{n0}}(e^{\pi kR}-1)}$ and the massive mode
spectrum is $n\pi k/(e^{\pi kR}-1)\;$ ($n=1,2,\cdots$).

We assume that the SM fields live on the 
infrared ($x^5=\pi R$) boundary where the fundamental scale is reduced
to TeV and the hierarchy problem is solved if $kR\sim10$~\cite{RS}. It
is a non-trivial task to obtain the analytic expression of
seesaw-induced masses by evaluating the mass matrix elements and
summing up the contributions of KK-mode integration. Here we consider
a simple and tractable case that Majorana masses are given only on 
the $x_5=0$ boundary~\cite{RS_braneM} (and bulk Dirac masses vanish,
just for simplicity), though the result given below does not depend on
whether the Majorana masses are placed on 
the $x^5=0$ or $x^5=\pi R$ boundary. Using the above wavefunctions, we
obtain the mass matrices in four-dimensional effective theory;
\begin{eqnarray}
  M_{K_{mn}} =\, -\frac{n}{R}h_n^2\delta_{mn}, \quad\;
  M_{R_{mn}} =\, \frac{h_mh_n}{\pi R}M, \quad\;\;
  M_{L_{mn}} =\, 0, \quad\;
  m_n \,=\, \frac{e^{\frac{1}{2}\pi kR}h_n}{\sqrt{\pi R}}\,m. \quad
\end{eqnarray}
One type of the Majorana masses, $M_L$, vanishes 
since $N_L^{\,n}$ have the negative $Z_2$ parity and the wavefunctions
become zero at the boundary. Another Majorana mass matrix, $M_R$, has
the off-diagonal ($m\neq n$) entries since the KK momentum is not
conserved at the boundary. The KK and Majorana masses receive the
exponential warp factors from the gravitational background.

It seems difficult to diagonalize the heavy-field mass 
matrix $M_N$ which is composed of $M_R$ and the KK 
masses $M_K$. However the mixing vertex $V$ between heavy modes and
the SM fields can be evaluated by taking the inverse of $M_N$ that is
given by
\begin{eqnarray}
  V \;=\; U_\nu^\dagger \big(%
  \begin{array}{cccccc}
  \!\tfrac{\sqrt{\pi R}}{-h_0}\,e^{\frac{1}{2}\pi kR}\,m^\dagger
  M^{-1} & \,0\, & \,0\, & \,0\, & \cdots
  \end{array}\big)U_N.
\end{eqnarray}
Notice that all the components but the 1st one are vanishing in the
interaction basis of heavy modes. From this mixing matrix, we obtain
the seesaw-induced masses and the heavy-mode mixing with the SM
neutrinos:
\begin{eqnarray}
  M_\nu &=& e^{\pi kR}\,m^{\rm t}M^{-1\,*}m, \\[1mm]
  \nu &\simeq& U_\nu\nu_d-\sqrt{\tfrac{2}{k}}\,e^{\pi kR}\,m^\dagger
  M^{-1} \epsilon N_R^{\,0\,*}.
\end{eqnarray}
It is found that the result is almost the same as in the standard
four-dimensional seesaw model with the heavy mass 
scale $M'=Me^{-\pi kR}\sim\text{TeV}$\@. Therefore the model on the
warped extra dimension cannot naively be made observable at collider
experiments because the heavy-mode mixing to the SM sector is roughly
given by $|V|\sim(M_\nu/M')^{1/2}\sim10^{-6}$ and is too small to be
detected. The conclusion would not be changed even if different curved
geometries are considered because the seesaw-induced mass is
determined without the knowledge of background metric~\cite{WY}.

An introduction of bulk Dirac masses modifies the wavefunctions of
bulk fermions $\Psi$. It is easily found that the Dirac masses lead to
additional wavefunction factors, which is similar to the result in
Section~\ref{sec:b_mass}, and hence does not cure the problem. Another
option is to extend the SM fields into the five-dimensional bulk and
to include their bulk Dirac masses. In low-energy effective theory,
the mixing elements $m_n$ between the SM neutrinos and bulk singlets
are suppressed if one chooses the bulk masses of left-handed leptons
such that they are localized away from the boundary at which the
neutrino Yukawa coupling is given. However even in this case, we could
not have the observable seesaw model in the sense that only the zero
mode is accessible and the higher KK-mode mixing with the SM neutrinos
is small.

Finally we comment on a possible modification of Model~2 given in
Section~\ref{sec:Model2} by considering the same field configuration
on the warped background. Unlike the flat background, the mass
eigenvalues of low-lying KK-excited modes are not dominated by the
bulk Dirac mass $m_d$ and their wavefunctions become un-suppressed. In
addition, the KK-excited modes are localized towards 
the $x_5=\pi R$ boundary and have stronger couplings to the SM
sector. These facts would make it possible to observe the right-handed
neutrinos at the LHC.

\bigskip\bigskip\bigskip

%%%%%%%%%%%%%%%%%%%%%%%%%%%%%%%%%%%%%%%%%%%%%%%%%%%%%%%%%%%%%%%%%
{\centering\section{Summary}%
\label{sec:summary}}

We have presented several seesaw scenarios in a five-dimensional
extension of the SM, where right-handed neutrinos live in the bulk and 
the SM particles stay at a four-dimensional boundary. The light
neutrino mass scale is of the order of eV, while the TeV-scale KK
neutrino modes have sizable gauge and Yukawa couplings to the SM
sector, which situation leads to observable signatures in future
particle experiments. We have discussed various extra-dimensional
schemes for making heavy states in the seesaw mechanism
observable. Among them, the collider signatures have been analyzed for
two illustrative models: the one involves the seesaw cancellation with
a particular value of bulk Majorana mass and another has light Dirac
neutrinos. Both models realize approximate lepton number conservation in
low-energy effective theory.

As the most effective LHC signal, we have analyzed the processes with
tri-lepton final 
states $pp\to\ell^\pm\ell^\pm\ell^\mp\nu(\bar\nu)$ and its
conjugates. We have extended our previous study to including three
types of neutrino mass patterns allowed by the current experimental
data. It is found that the scenarios give excessive tri-lepton events
beyond the SM background in wide regions of parameter space and the
LHC would discover a sign of tiny neutrino mass generation. Further,
as for the three-generation mixing, the cross sections are controlled
by the MNS neutrino mixing matrix, and therefore a detailed
measurement of branching ratios would corroborate the lepton flavor
structure and the Type I seesaw scheme, which is left for future study.

In the present analysis, the signal essentially receives the
contribution only from the 1st excited mode and is difficult to
discriminate the seesaw mechanism in higher dimensions from other
models for neutrino mass generation. The observation of higher KK
modes is expected to be within the reach of future particle
experiments such as the ILC\@. That could substantially confirm the
existence of extra spatial dimensions in Nature.

\bigskip\bigskip\bigskip
%%%%%%%%%%%%%%%%%%%%%%%%%%%%%%%%%%%%%%%%%%%%%%%%%%%%%%%%%%%%%%%%%
{\centering\subsection*{Acknowledgments}}
\noindent
This work is supported by the scientific grants from the ministry of
education, science, sports, and culture of Japan (No.~20740135,
21740174, 22244021), and also by the grant-in-aid for the global COE
program "The next generation of physics, spun from universality and
emergence" and the grant-in-aid for the scientific research on
priority area (\#441) "Progress in elementary particle physics of the
21st century through discoveries of Higgs boson and supersymmetry"
(No.~16081209).

\newpage
%%%%%%%%%%%%%%%%%%%%%%%%%%%%%%%%%%%%%%%%%%%%%%%%%%%%%%%%%%%%%%%%%

\end{document}